\documentclass{article}
\usepackage{amsmath}
\usepackage{appendix}

\input{tcilatex}
\begin{document}

\title{Probing new physics through $B\rightarrow Kl^{+}l^{-}$ decays in
R-parity violating minimal supersymmetric standard model}
\author{\ \ \ $Azeem\ Mir\footnotemark $,$\ Farida\ Tahir\footnotemark $, $%
Kamaluddin\ Ahmed\footnotemark \ \ \ $ \\
%EndAName
(\textit{Physics Department, COMSATS Institute of Information Technology,
Islamabad.})}
\date{}
\maketitle

\begin{abstract}
\textit{We study the decay rate of process }$B\rightarrow K\
l^{+}l^{-}(l=e,\mu )$ and some of its other related observables, like
forward backward asymmetry ($A_{FB}$)\ ,polarization asymmetry ($PA$) and
CP-asymmetry ($A_{CP}$) in\textit{\ R-parity violating (}$\NEG{R}_{p}$%
\textit{) Minimal Supersymmetric Standard Model (MSSM). The analysis shows
that} $\NEG{R}_{p}$\textit{Yukawa coupling products contribute significantly
to }the branching fraction\ of $B\rightarrow K\ l^{+}l^{-}$ within 1$\sigma $
and 2$\sigma $\textit{. Study shows that }$PA$\textit{\ and }$A_{FB}\ $are
sensitive enough to $\NEG{R}_{p}$\textit{Yukawa coupling products and turn
out to be good predictions for measurement in future experiments}$.$The
CP-asymmetry calculated in this framework agrees well with the recently
reported value(i.e. 7\%).
\end{abstract}

\section{Introduction}

$B$ physics has played a central role in clarifying issues related to the
standard model (SM) like measurements of Cabibbo-Kobayashi-Maskawa (CKM)
unitary angles, probing CP-violation and B decays sensitive to new physics%
\cite{BELLE}. Belle and BaBar experiments have contributed significantly to
this important area of particle physics which holds promise and potential to
discover physics beyond the standard model or new physics (NP) and offers a
motivation for construction of super B-factories \cite{SB}. Various authors
have studied the contribution of NP, like 2 Higgs doublet Model$(2HDM),MSSM$
and extra dimensions, etc, to $B$ decays\cite{NP}. There are some important
processes in B-physics interesting to probe for NP involving $B\overline{B}$
mixing and leptonic B-decays \cite{BB}. Leptonic $B$ decays are simple to
analyze and may provide some clear signals for exploring physics beyond SM.
Among such processes are the FCNC (flavour changing neutral current)
processes. FCNC proceed through higher order box and penguin diagrams in SM,
whose contribution may compete with that of leading contributions of
R-parity violating MSSM. While the branching fraction in such processes may
be an obvious choice for testing bounds on parameters related to physics
beyond the SM or NP, different observables like $A_{FB},\ PA$ and $CP$
asymmetry may also show sensitivity to NP parameters. Several authors have
studied NP contribution to these observables in different models like $2HDM$
and extra dimensions and confirmed that these observables are indeed
sensitive to parameters of NP\cite{TM,OP}.

Supersymmetry(SUSY) was proposed to extend the Standard Model and to define
a more complete theoretical framework which will unify all interactions
including gravity. SUSY among other issues solves the gauge hierarchy
problem related to Higgs boson and leads to the hint for unification of
forces\cite{SU}. As an extended symmetry of space-time, it gives rise to
spin doubling of the constituent and force fields and defines a generalized
superpotential which contains a part conserving R-parity and another
allowing R-parity violation\cite{H1}. SUSY\ with finite neutrino Majorana
mass leads to lepton flavor mixing and oscillations, an important area of
recent research. Supersymmetric partners(spartners) have, however, not been
found in nature yet, though the search for SUSY is now focussed on LHC with
expectations. SUSY, however, remains as a promising candidate theory and has
made its way into many other NP scenarios like extra dimensions and strings 
\cite{SP}.

One of the drawbacks of SUSY is that matter is no longer predicted to be
stable as proton can decay through sparticles. In order to avoid this
catastrophic scenario, R-parity was introduced\cite{H1}. R-parity is a
discrete symmetry and is defined as $R=(-1)^{3B+L+2S}$, where $B,L$ and $S$
are baryon ,lepton number, and spin of a particle respectively. This implies
that sparticles cannot mediate an interaction and the sparticles are always
produced in pair. The second consequence implies the existence of a lightest
supersymmetric sparticle, which is a very good candidate for dark matter.
R-parity, though, ensures the stability of matter, is an ad hoc assumption.
R-parity violation, however, cannot be ruled out theoretically. R-parity
violation ($\NEG{R}_{p}$), is introduced by relaxation of R-parity
conservation with some constraints on couplings responsible for proton
decay. $\ $

The $\NEG{R}_{p}$ superpotential approach results in a framework which is
phenomenologically rich and has been actively pursued. Lepton flavor and
number violating decays are mediated by sparticles. In addition, neutrino
also acquires mass\cite{MY}. FCNC constitutes as one important part of $\NEG%
{R}_{p}$SUSY phenomenology. The aim of this paper is to discuss how well the
experimentally observed value of branching fraction of FCNC ($b\rightarrow
sl^{+}l^{-}$)fits with the theoretical predictions on the basis of $\NEG%
{R}_{p}$. Further, using SUSY, to examine, parameters which fit the decay
rate in the process to explain \ forward backward, polarization, and CP-
asymmetries in this decay for comparison with future expected experiments.

In Sec. 2, we review the effective Hamiltonian and also $\NEG{R}_{p}$ $SUSY\ 
$potential approach for use in our calculation. In Sec.3 we review and
analyze some of the results on form factors from existing literature to be
used in our study based on R$_{p}$violating MSSM. Further, in this section
we give the calculation, in this framework, for the decay rate of the decay
processes $B\rightarrow Kl^{+}l^{-},l=e,\mu ,$and$\ $the relevant various
asymmetries to this process.In Sec.4, we give results and implications of
our results from the point of view of possibile comparison of measured
values in future experiments. Last section Sec. 5, gives conclusion and
summarizes our results.

\section{$B\rightarrow Kl^{-}l^{+}$ decay in SM and $\NEG{R}_{p}$ SUSY}

The effective Hamiltonian for the given decay process is given by\cite{SM,TM}

\begin{equation}
H_{eff}=\frac{4G_{F}}{\sqrt{2}}V_{tb}V_{ts}^{\ast }\underset{i=1}{\overset{10%
}{\Sigma }}\{C_{i}(\mu )\ O_{i}(\mu )+C_{Q_{i}}(\mu )\ O_{Q_{i}}(\mu )\}.
\end{equation}

The first set of operators in eq. (1) describe the contribution from SM,
while the second term describes the contribution from physics beyond SM. $%
\mu $ defines the energy scale of interactions. The corresponding Wilson
coefficients for SM can be found in literature\cite{TM,SM}. Form factors are
needed to calculate the decay rate of exclusive $B\rightarrow
Kl^{+}l^{-};l=e,\mu \ $channel. We use the form factors analytically derived
in literature\cite{SM}.

Since we are interested in comparing the SUSY contribution along with that
of SM in the observed value of branching fraction, we proceed to consider
MSSM($N=1$) super potential \cite{SW,H1}\smallskip 
\begin{equation}
\ W_{\text{\textsl{$\;\NEG{R}$}}_{\text{\textbf{p}}}}=\frac{1}{2}\lambda
_{ijk}L_{i}L_{j}E_{k}^{c}+\lambda _{ijk}^{^{\prime }}L_{i}Q_{j}D_{k}^{c}+%
\frac{1}{2}\lambda _{ijk}^{^{\prime \prime
}}U_{i}^{^{C}}D_{j}^{^{C}}D_{k}^{^{C}}+\mu _{i}H_{u}L_{i},\ \ \ \ \ \ \ \ \
\smallskip
\end{equation}%
where $i,\,j,\,k$\ are generation indices, $L_{i}$\ and $Q_{i}$\ are the
lepton and quark left-handed $SU(2)_{L}$\ doublets and $E^{c}$, $D^{c}$\ are
the charge conjugates of the right-handed leptons and quark singlets,
respectively. $\lambda _{ijk}$\ , $\lambda _{ijk}^{\prime }$\ \ and \ $%
\lambda _{ijk}^{\prime \prime }$ are Yukawa couplings. Note that the term
proportional to $\lambda _{ijk}$\ is antisymmetric in the first two indices $%
[i,\,j]$\ and $\lambda _{ijk}^{\prime \prime }$ is antisymmetric in the\ last%
$\ $two indices $[j,k]$, implying $9(\lambda _{ijk})+27(\lambda
_{ijk}^{^{\prime }})+9\left( \lambda _{ijk}^{^{\prime \prime }}\right) =45$\
independent coupling constants among which 36 are related to the lepton
flavor violation (9 from $LLE^{c}$\ and 27 from $LQD^{c}$). The last term
can be rotated away by a unitary transformation. However this may induce
additional terms involving Yukawa couplings, not relevant here \cite%
{R.Barbie}. In $\NEG{R}_{p}$ MSSM the relevant effective Lagrangian is given
by \cite{H1}\smallskip

\begin{equation}
L_{\QTR{sl}{\NEG{R}\;}_{\text{\textbf{p}}}}^{eff}\left( \ \overline{b}%
\longrightarrow \bar{s}\ l_{\beta }+\overline{l}_{\beta }\right) =\frac{G_{F}%
}{2\sqrt{2}\pi }\left[ 
\begin{array}{c}
A_{\beta \beta }^{bs}\left( \overline{l_{\beta }}\gamma ^{\mu }P_{L}l_{\beta
}\right) \left( \overline{s}\gamma _{\mu }P_{R}b\right) \\ 
-B_{\beta \beta }^{bs}\left( \overline{l_{\beta }}P_{R}l_{\beta }\right)
\left( \overline{s}P_{L}b\right) \\ 
-C_{\beta \beta }^{bs}\left( \overline{l_{\beta }}P_{L}l_{\beta }\right)
\left( \overline{s}P_{R}b\right)%
\end{array}%
\right] ,\smallskip \beta =e,\mu
\end{equation}%
where $P_{R}=\frac{1+\gamma _{5}}{2};\ P_{L}=\frac{1-\gamma _{5}}{2}.$The
first term in eq.(3) comes from the up squark exchange and the remaining two
terms come from sneutrino exchange. Here $s$ and $b$\ denote strange and
beauty down type quarks. The dimensionless coupling constants $A_{\beta
\beta }^{bs},$ $B_{\beta \beta }^{bs}\ $and $C_{\beta \beta }^{bs}\ $depend
on the species of charged leptons and are given by \cite{HK}\smallskip 
\begin{equation}
A_{\beta \beta }^{bs}=\frac{2\sqrt{2}\pi }{G_{F}}\underset{m,n,i=1}{\overset{%
3}{\sum }}\frac{V_{ni}^{\dagger }V_{im}}{2m_{\widetilde{u_{i}^{c}}}^{2}}%
\lambda _{\beta nk}^{\prime }\lambda _{\beta mp}^{\prime \ast },
\end{equation}%
\begin{equation}
B_{\beta \beta }^{bs}=\frac{2\sqrt{2}\pi }{G_{F}}\underset{i=1}{\overset{3}{%
\sum }}\frac{2}{m_{\widetilde{\nu }_{Li}}^{2}}\lambda _{i\beta \beta }^{\ast
}\lambda _{ipk}^{\prime },
\end{equation}%
\begin{equation}
C_{\beta \beta }^{bs}=\frac{2\sqrt{2}\pi }{G_{F}}\underset{i=1}{\overset{3}{%
\sum }}\frac{2}{m_{\widetilde{\nu }_{Li}}^{2}}\lambda _{i\beta \beta
}\lambda _{ikp}^{\prime \ast }.\smallskip
\end{equation}

In the Section 3 we study the asymmetries relevant to this process using the
above described R-parity violating MSSM framework which essentially
involves, in this problem, simple computation of Feynman diagrams with
exchange of SUSY particles(Fig.1).

\section{Calculations and Analysis}

The matrix element of the decay rate ($B\rightarrow Kl^{+}l^{-}$) is given by%
\cite{SM}

\begin{equation}
M=F_{S}\bar{l}l+F_{V}p_{\mu }\bar{l}\gamma ^{\mu }l+F_{A}(p_{B})_{\mu }\bar{l%
}\gamma ^{\mu }\gamma ^{5}l+F_{P}\bar{l}\gamma ^{5}l
\end{equation}

where $(p_{B})_{\mu }$ is the initial momentum of $B$ meson and $F_{S,V,A,P}$
are functions of lorentz invariant quantities like dilepton centre of mass
energy squared($s$). These functions involve the Wilson coefficients ($%
C_{i}(\mu )$ and $C_{Q_{i}}(\mu )$) as given below in eq.(11) are reproduced
from \cite{SM}.

The Forward-Backward Asymmetry ($A_{FB}$) is related to asymmetric angular
distribution of dilepton pair with repect\ to the initial meson direction of
momentum in the dilepton rest frame. The $A_{FB}$ is defined as:%
\begin{equation}
A_{FB}(s)=\frac{\int_{0}^{1}d\cos \theta \frac{d^{2}\Gamma }{dsd\cos \theta }%
-\int_{-1}^{0}d\cos \theta \frac{d^{2}\Gamma }{dsd\cos \theta }}{%
\int_{0}^{1}d\cos \theta \frac{d^{2}\Gamma }{dsd\cos \theta }%
+\int_{-1}^{0}d\cos \theta \frac{d^{2}\Gamma }{dsd\cos \theta }}.
\end{equation}

In the dilepton rest frame \cite{CB,CH}

\begin{equation}
A_{FB}(s)=\frac{1}{128\pi ^{3}m_{B}^{3}(\frac{d\Gamma }{ds})}m_{l}\beta
(m_{l},s)^{2}\lambda (s)\func{Re}(F_{S}F_{V}^{\ast }),
\end{equation}%
where

\begin{equation}
\frac{d\Gamma }{ds}=\frac{1}{256\pi ^{3}m_{B}^{3}}\beta (m_{l},s)\lambda ^{%
\frac{1}{2}}(s)R(s)
\end{equation}

\begin{center}
$\beta (m_{l},s)=(1-\frac{4m_{l}^{2}}{s})^{\frac{1}{2}}$\bigskip ,\ 

$\lambda
(s)=m_{B}^{4}+m_{K}^{4}+s^{2}-2sm_{K}^{2}-2sm_{B}^{2}-2m_{K}^{2}m_{B}^{2},%
\medskip $
\end{center}

where

\begin{eqnarray*}
R(s) &=&\left\vert F_{S}\right\vert ^{2}2s\beta (m_{l},s)^{2}+\left\vert
F_{P}\right\vert ^{2}2s+\left\vert F_{V}\right\vert ^{2}\frac{1}{3}\lambda
(s)(1+\frac{2m_{l}^{2}}{s})+\left\vert F_{A}\right\vert ^{2}[\frac{1}{3}%
\lambda (s)(1+\frac{2m_{l}^{2}}{s})+8m_{B}^{2}m_{l}^{2}] \\
&&+\func{Re}(F_{P}F_{A}^{\ast })4m_{l}(m_{B}^{2}-m_{K}^{2}+s),
\end{eqnarray*}

and $s$ (invariant mass squared of the dilepton system in its rest frame) is
bounded as:

\begin{center}
$4(m_{l})^{2}\leq s\leq (m_{B}-m_{K})^{2}$
\end{center}

The coefficients $F_{V},F_{A},F_{S},F_{P}$ are determined by using eqs.(1, 7
\& A-1 to A-4)

\begin{eqnarray}
F_{V} &=&\frac{G_{F}\alpha V_{tb}V_{ts}^{\ast }}{2\sqrt{2}\pi }%
(2C_{9}^{eff}f^{+}(s)-C_{7}\frac{4m_{b}}{m_{B}+m_{K}}f^{T}(s))+\frac{1}{4}%
f^{+}(s)\underset{i,m,n=1}{\overset{3}{\sum }}V_{ni}^{\dagger }V_{im}\frac{%
\lambda _{\beta n3}^{\prime }\lambda _{\beta m2}^{\prime \ast }}{m_{%
\widetilde{u_{i}^{c}}}^{2}}, \\
F_{A} &=&\frac{G_{F}\alpha V_{tb}V_{ts}^{\ast }}{2\sqrt{2}\pi }%
(2C_{10}f^{+}(s))-\frac{1}{4}f^{+}(s)\overset{3}{\underset{i,m,n=1}{\sum }}%
V_{ni}^{\dagger }V_{im}\frac{\lambda _{\beta n3}^{\prime }\lambda _{\beta
m2}^{\prime \ast }}{m_{\widetilde{u_{i}^{c}}}^{2}},  \notag \\
F_{S} &=&\frac{1}{4}\frac{(m_{B}^{2}-m_{K}^{2})}{m_{b}-m_{s}}f_{o}(s)%
\underset{i=1}{\overset{3}{\sum }}\frac{(\lambda _{i\beta \beta }^{\ast
}\lambda _{i23}^{\prime }-\lambda _{i\beta \beta }\lambda _{i32}^{\prime
\ast })}{m_{\widetilde{\nu }_{Li}}^{2}},  \notag \\
F_{P} &=&\frac{G_{F}\alpha V_{tb}V_{ts}^{\ast }}{2\sqrt{2}\pi }%
2m_{l}C_{10}(f^{+}(s)+f^{-}(s))+\frac{1}{4}\frac{(m_{B}^{2}-m_{K}^{2})}{%
m_{b}-m_{s}}f_{o}(s)\underset{i=1}{\overset{3}{\sum }}\frac{(\lambda
_{i\beta \beta }^{\ast }\lambda _{i23}^{\prime }+\lambda _{i\beta \beta
}\lambda _{i32}^{\prime \ast })}{m_{\widetilde{\nu }_{Li}}^{2}},  \notag
\end{eqnarray}

Notice that in eq. (11), $F_{V},F_{A},F_{P},$ under $H_{eff}$ (see eq.(1))
are split-up into two parts each carrying SM\&MSSM contributions, while $%
F_{S}$ contains only the contribution from MSSM.

Also here, $f^{\pm }(s),f^{T}(s),f_{o}(s)$ are form factors\cite{SM}related
to decay $B\rightarrow Kl^{+}l^{-}$. We use values of Wilson coefficients $%
C_{i}(m_{b})\ \cite{OP,SM,CB}$ to leading order evaluated at the rest mass $%
m_{b}$ of b quarks (see eqs. A-5 to A-7 of Appendix A). Inegrated branching
fracton for this process used in our calculation for comparison with the
data on branching ratio is given

\begin{equation}
BR(B^{\pm }\rightarrow K^{\pm
}l^{+}l^{-})=\int_{4(m_{l})^{2}}^{(m_{B}-m_{K})^{2}}\frac{d\Gamma }{ds}ds\
\tau _{o}
\end{equation}

where $\tau _{o}$ is the life time of $B$ meson.

The polarization asymmetries are defined, as usual, as \cite{OP,SM}\medskip :

\begin{equation}
(PA)_{i}=\frac{\frac{d\Gamma }{ds}(\widehat{n}=-\widehat{e_{i}})-\frac{%
d\Gamma }{ds}(\widehat{n}=\widehat{e_{i}})}{\frac{d\Gamma }{ds}(\widehat{n}=-%
\widehat{e_{i}})+\frac{d\Gamma }{ds}(\widehat{n}=\widehat{e_{i}})},
\end{equation}

where ($i=L,N,T$) and$\ \widehat{n}$ is the spin direction of lepton$\ l$.
Following three polarization unit vectors are defined in the centre of mass
of the $l^{+}l^{-}$ system and are given in literature\cite{OP,SM}:\medskip

\begin{center}
$\widehat{e_{L}}=\frac{\vec{p}_{1}}{\left\vert \vec{p}_{1}\right\vert };$

$\widehat{e_{N}}=\frac{\vec{p}_{K}\times \vec{p}_{1}}{\left\vert \vec{p}%
_{K}\times \vec{p}_{1}\right\vert };$

$\widehat{e_{T}}=\widehat{e_{N}}\times \widehat{e_{L}};\medskip $
\end{center}

where $\vec{p}_{1}$ and $\vec{p}_{K}$ are the three momenta of the $l^{-}$
lepton and the $K$ meson respectively. $\widehat{e_{L}}$ is the unit vector
of lepton in the dilepton rest frame of reference. Unit vector $\widehat{%
e_{N}}$ is normal to the plane of $K$ meson and $\widehat{e}_{L}$ and $%
\widehat{e_{T}}$ lie in the plane containing lepton and $K$ meson momentum
vector.

We plot ($PA)_{L}$and$(PA)_{T}$ versus dilepton centre of mass energy and
Yukawa couplings to gain an insight into the behavior of these observables
in physics beyond SM(based on MSSM). Normal polarization asymmetry($\widehat{%
e_{N}}$) is too small to be given any importance\cite{OP}. The relevant
expressions for polarization asymmetries as taken from \cite{TM,OP,SM}

\begin{equation}
(PA)_{L}=\frac{\beta (m_{l},s)}{R(s)}[\frac{2}{3}\lambda \func{Re}%
(F_{V}^{\ast }F_{A})-4s\func{Re}(F_{S}^{\ast
}F_{P})-4m_{l}((m_{B})^{2}-(m_{K})^{2}-s)\func{Re}(F_{A}^{\ast }F_{S})],
\end{equation}

\begin{equation}
(PA)_{T}=\frac{\pi \sqrt{\lambda (s)}}{R(s)\sqrt{s}}%
[m_{l}((m_{B})^{2}-(m_{K})^{2}-s)\func{Re}(F_{V}^{\ast }F_{A})+s(\beta
(m_{l},s))^{2}\func{Re}(F_{A}^{\ast }F_{S})+s\func{Re}(F_{V}^{\ast }F_{P})],
\end{equation}

where $m_{l}$ is the mass of lepton

We have also computed average polarization asymmetries, which are defined as

\begin{equation}
<(PA)_{i}>=\frac{\int_{4(m_{l})^{2}}^{(m_{B}-m_{K})^{2}}(PA)_{i}\frac{dBr}{ds%
}ds}{\int_{4(m_{l})^{2}}^{(m_{B}-m_{K})^{2}}\frac{dBr}{ds}ds}\medskip ,
\end{equation}%
Number of events required to calculate $PA$ at $n\sigma $ level are given by$%
\ N=n^{2}/(BR)(<(PA)_{i}>)^{2}$, where $n$ is the desired $(1\sigma $ $or\
2\sigma )$ level. $BR$ represents the branching fraction of $B\rightarrow Kl%
\bar{l}$ and $<(PA)_{i}>$ is the polarization asymmetry. CP asymmetries are
defined as\cite{CP}

\begin{equation}
A_{CP}=\frac{\frac{d\Gamma (B\rightarrow Kl^{+}l^{-})}{ds}-\frac{d\Gamma (%
\bar{B}\rightarrow \bar{K}l^{-}l^{+})}{ds}}{\frac{d\Gamma (B\rightarrow
Kl^{+}l^{-})}{ds}+\frac{d\Gamma (\bar{B}\rightarrow \bar{K}l^{-}l^{+})}{ds}}
\end{equation}

In SM, $C_{9}^{eff}$ becomes complex due to non-negligible terms induced in $%
A\&B$\cite{CP1}. Its complex nature is responsible for CP-asymmetry. $%
C_{7}\&C_{10}$ being real do not contribute to CP-asymmetry. $\NEG{R}_{p}$
Yukawa couplings can be imaginary so we have(see Appendix-B )

\begin{eqnarray}
\frac{d\Gamma (\bar{B}\rightarrow \bar{K}l^{-}l^{+})}{ds} &=&\frac{1}{256\pi
^{3}m_{B}^{3}}\beta (m_{l},s)\lambda ^{\frac{1}{2}}(s)\{\left\vert \bar{F}%
_{S}\right\vert ^{2}2s\beta (m_{l},s)^{2}+\left\vert \bar{F}_{P}\right\vert
^{2}2s+\left\vert \bar{F}_{V}\right\vert ^{2}\frac{1}{3}\lambda (s)(1+\frac{%
2m_{l}^{2}}{s})  \notag \\
&&+\left\vert \bar{F}_{A}\right\vert ^{2}[\frac{1}{3}\lambda (s)(1+\frac{%
2m_{l}^{2}}{s})+8m_{B}^{2}m_{l}^{2}]+\func{Re}(\bar{F}_{P}^{\ast }\bar{F}%
_{A})4m_{l}(m_{B}^{2}-m_{K}^{2}+s)\},
\end{eqnarray}

\begin{equation}
A_{CP}=\lambda (s)(1+\frac{2m_{l}^{2}}{s})(\left\vert F_{V}\right\vert
^{2}-\left\vert \bar{F}_{V}\right\vert ^{2})(D(s))^{-1}
\end{equation}

where

\begin{eqnarray}
D(s) &=&6(\left\vert F_{S}\right\vert ^{2}2s\beta (m_{l},s)^{2}+\left\vert
F_{P}\right\vert ^{2}2s+\left\vert F_{A}\right\vert ^{2}[\frac{1}{3}\lambda
(s)(1+\frac{2m_{l}^{2}}{s})+8m_{B}^{2}m]+ \\
&&\func{Re}(F_{P}F_{A}^{\ast })4m_{l}(m_{B}^{2}-m_{K}^{2}+s)+\frac{\lambda
(s)}{6}(1+\frac{2m_{l}^{2}}{s})(\left\vert F_{V}\right\vert ^{2}+\left\vert 
\bar{F}_{V}\right\vert ^{2}))  \notag
\end{eqnarray}

One can easily see from the expression of $F_{V}\ \&\bar{F}_{V}$ from
appendix B that if SUSY terms containing $\lambda _{\beta n3}^{\prime
}\lambda _{\beta n2}^{\prime \ast }$ vanish in eq.(20), then one recovers SM
result for $A_{CP}$.

Also it is clear from the above expression that $A_{CP}$ directly depends
upon the squark exchange $\NEG{R}_{p}\ $coupling(see Appendix-B). We have
calculated average CP-asymmetries to determine its dependence on $\NEG%
{R}_{p} $ Yukawa coupling products.

In the next Section, we give some results on asymmetries in this process
based on computations of the eqs (8-19) and discuss possibility of their
measurement in future experiments.

\section{Results and Discussion}

Fig. 1 gives the tree s-particle exchange Feynman diagrams evaluated as
leading SUSY contributions to this process. We have used data from pdg\cite%
{pdg} to plot Figs (2-10). The total branching ratio of decay processes $%
(B^{\pm }\rightarrow K^{\pm }l^{+}l^{-})$ are numerically calculated to be $%
5.17\times 10^{-7}$ for $l=e$ and $5.13\times 10^{-7}$ for $l=\mu ($see
eq.(12))$.$ The small difference between the two modes is due to difference
in phase space integration.

We have used bounds on $\NEG{R}_{p}$ couplings from the literature\cite{HK}.
In Figs. (2-3), integrated branching fraction has been plotted against $\NEG%
{R}_{p}$ Yukawa couplings within 1$\sigma $ (dashed) and 2$\sigma $(solid)
levels in observed value of branching fraction in the process $B^{\pm
}\rightarrow K^{\pm }l^{+}l^{-}$. Single coupling dominance has been assumed
in these graphs to observe how significant is the contribution of individual
couplings to the branching fraction. This is a clear demonstration of the
fact that tree level graphs in $\NEG{R}_{p}$ have a significant contribution
to FCNC. An interesting conclusion that can be drawn from these graphs
Figs(2-3) is that sneutrino $\NEG{R}_{p}$ couplings interfere only
constructively with the SM box and penguin graphs i.e. the branching
fraction in Fig (2) is enhanced. While the squark $\NEG{R}_{p}$ couplings
interfere destructively i.e. the branching fraction in Fig (3) can be
smaller than the SM prediction. This is because of a positive contribution
in $F_{V}$ and a negative contribution in $F_{A}$ (see eq.11). Figs (2-3)
also constrain the bounds on squark and sneutrino $\NEG{R}_{p}$ couplings.
Fig. (4) represents combined effect of both squark and sneutrino $\NEG%
{R}_{p} $ couplings. We have ignored the effects of $\lambda _{323}^{\prime
}\lambda _{311}$ as it only changes the given graphs slightly. We also
ignore the phase of $\lambda _{323}^{\prime \ast }\lambda _{3\beta \beta
}^{^{\prime }}(\beta =1,2)\ $in these graphs for a simple analysis.The
significant contribution made by these $\NEG{R}_{p}$ Yukawa couplings
indicate that one must look for more events of $(B^{\pm }\rightarrow K^{\pm
}\mu ^{+}\mu ^{-})$ in $LHCb$ and Super $B$ factory.

The decay process $B^{\pm }\rightarrow K^{\pm }\mu ^{+}\mu ^{-}$ is ideal to
measure new physics observables like $A_{FB},\ (PA)_{L}\&(PA)_{T}.$ Firstly
this decay has been observed experimentally\cite{Obs} and secondly these
observables are no longer suppressed in the decay process for muons as they
are in the decay process involving electrons and positrons due to relatively
large mass of muons. In Figs.(5-7), $A_{FB},\ (PA)_{L}$and$(PA)_{T}$ have
been plotted against dilepton centre of mass energy squared and different
set of $\NEG{R}_{p}$ Yukawa couplings. We have ignored the effects of phase
of $\lambda _{323}^{\prime \ast }\lambda _{3\beta \beta }^{^{\prime }}(\beta
=1,2)\ $in these graphs for a simple analysis.$\ A_{FB}$ is found to lie
between (-5 and +5)\%. It may be measured in experiments expected to take
place at Super B factory \cite{SB}. $(PA)_{L}$ in SM is negative at high
dilepton centre of mass energy values but in the presence of $\NEG{R}_{p}$
Yukawa couplings it also changes sign and can become as large as 50\% as
shown in Fig. (6). Thus a change of sign of $(PA)_{L}$ will be a strong
indication of new physics beyond SM. The magnitude of $(PA)_{T}$ is
(10-60)\% in SM, but its magnitude may rise upto 90\% in the presence of $%
\NEG{R}_{p}$ Yukawa couplings as shown in Fig. (7). Thus $PAs$ are indeed
sensitive to $\NEG{R}_{p}$ Yukawa couplings and are a better probe to NP
than $A_{FB}$.

We next calculate average $PA$. For $B^{\pm }\rightarrow K^{\pm }\mu ^{+}\mu
^{-}$, we find that $<(PA)_{T}>_{SM}=-0.11$ and $<(PA)_{L}>_{SM}=-0.49$.
Fig(8) shows that average $PAs$ change significantly in the presence of $\NEG%
{R}_{p}$ Yukawa couplings. Calculations further show an estimate of $6\times
10^{6}n^{2}\ B\bar{B}$ events are required for the measurement of $(PA)_{T}$%
, while $1\times 10^{7}n^{2}$ events are required for the measurement of $%
(PA)_{L}$, where $n$ is $1\sigma $ or $2\sigma $ level. We have also
calculated average $A_{FB}$ in Fig(9). It is found to be ($\pm 4$)\%.i.e. it
requires an order of $(10^{9}-10^{11})n^{2}B\bar{B}$ events. In Super B
factories there is an estimate of $10^{9}B\bar{B}$ available events. Thus
the transverse polarization asymmetry in this process appears to be a
dominant observable for looking beyond SM in a framework based on MSSM. We,
thus look forward for measuring $PA^{^{\prime }}s$ and $A_{FB}$ in Super B
factories for comparison with asymmetry predictions made in this work. In
the following we argue that CP-asymmetry is another significant and
important observable for looking for physics beyond standard model in such
processes.

In Fig (10a), we show the behavior of CP-asymemtry in SM. Its average value
is -$1\times 10^{-3},$ whose magnitude is quite small as compared to the
experimentally measured value i.e. $0.07\cite{pdg}.$ Fig (10(b)) shows the
behavior of CP-asymmetry in $\NEG{R}_{p}$\ MSSM. Since the contribution by
the squark exchange term is proportional to $G_{F}\alpha V_{tb}^{\ast
}V_{ts} $, whereas the pure SM contribtion to $A_{CP}$ is proportional to $%
(G_{F}\alpha V_{tb}^{\ast }V_{ts})^{2}($see appendix B$).$It shows that $CP$
asymmetry is influenced by the imaginary value of squark exchange term
significantly as compared to the SM contribution. We study CP-asymmetry in
the abscence of sneutrino exchange term because it is not much influenced by
them. The maximum contribution by imaginary part of squark exchange term is
fairly close to the experimental value, i.e 7\%, although the experimental
errors are rather large (see Fig. 10b). Also $CP$ asymmetry beomes
vanishingly small when the squark exchange term is real (i.e. imaginary part
is zero in eq.(19)). We believe that better statistics will be available in
Super B factories such that the CP asymmetries in such processes will be
measured with better precision to match these predictions.

\section{Conclusion \& Summary}

We have carried out an analysis of the decay processes $(B^{\pm }\rightarrow
K^{\pm }l^{+}l^{-})$ in $\NEG{R}_{p}$ SUSY. The analysis shows that the
contribution of $\NEG{R}_{p}$ Yukawa couplings to branching fraction of this
decay along with that of SM is significant. This makes $\NEG{R}_{p}$ SUSY a
good framework to study FCNC semileptonic decays. The new physics
observables including asymmetries($A_{FB}$,$PAs$ and $A_{CP})$ have also
been calculated. Sneutrino $\NEG{R}_{p}$ Yukawa couplings play an important
role in these observables by providing different kinds of enhancements. Thus
a non zero $<A_{FB}>$ and a relatively high magnitude$<(PA)_{L}>\ $and$%
<(PA)_{T}>$ will indicate direct signs of new physics. Transverse
polarization asymmetry is one of the most favorable physics observable in
these searches as it is negligible in SM but rises to a significant
magnitude in the presence of $\NEG{R}_{p}$ Yukawa couplings. In addition, $%
\NEG{R}_{p}$ Yukawa squark couplings also have sizeable contribution to
CP-asymemtry. These observables may be measured at future experiments at
Super B factory\cite{SB} and may provide a good probe for physics beyond
Standard Model.

\subsection{\textbf{Acknowledgement}}

Azeem\ Mir is indebted to the Higher Education Commission of Pakistan for
financial support.

\newpage

\begin{center}
\textbf{Appendix-A}

\textbf{Wison Coefficients}
\end{center}

We reproduce the matrix element of the decay rate ($B\rightarrow Kl^{+}l^{-}$%
) from\cite{SM}

\begin{equation*}
M=F_{S}\bar{l}l+F_{V}p_{\mu }\bar{l}\gamma ^{\mu }l+F_{A}(p_{B})_{\mu }\bar{l%
}\gamma ^{\mu }\gamma ^{5}l+F_{P}\bar{l}\gamma ^{5}l
\end{equation*}%
\bigskip

The hadronic matrix elements contained in $F_{V,A,S,P}$are given in terms of
form factors $f^{\pm }(s),f^{T}(s)$\cite{OP,SM,CB}$\medskip $

\begin{center}
\begin{equation}
<K(p_{K})|\bar{s}\gamma _{\mu }(1-\gamma _{5})b|B(p_{B})>=(p_{B}+p_{K})_{\mu
}f^{+}(s)+p_{\mu }f^{-}(s)\   \tag{A-1}
\end{equation}%
\ \ \ \ \ \ \ \ \ \ \ \ \ \ \ \ \ \ \ \ \ \ \ \ \ \ \ \ \ \ \ \ \ \ \ \ \ \
\ \ 

\begin{equation}
<K(p_{K})|\bar{s}i\sigma _{\mu v}\gamma ^{v}(1+\gamma
_{5})b|B(p_{B})>=(p_{B}+p_{K})_{\mu }s-p_{\mu }(m_{B}^{2}-m_{K}^{2})\frac{%
f^{T}(s)}{m_{B}+m_{K}},  \tag{A-2}
\end{equation}%
$\medskip $%
\begin{equation}
<K(p_{K})|\bar{s}b|B(p_{B})>=\frac{(m_{B}^{2}-m_{K}^{2})}{m_{b}-m_{s}}%
f_{o}(s),  \tag{A-3}
\end{equation}%
$\medskip $
\end{center}

where $p_{\mu }=(p_{B}-p_{K})_{\mu }$ is the momentum transfer to dilepton
pair and $s=p_{\mu }p^{\mu },m_{B},m_{K},m_{b},m_{s}$ are masses of B and K
mesons, b quark and s quark respectively$.$ We have computed form factors $%
f^{\pm }(s),f^{T}(s)$ from the analytic form given by\cite{SM}, which are
used in our plots.$f_{o}(s)$ is defined as

\begin{equation}
f_{o}(s)=f^{+}(s)+\frac{s}{m_{B}^{2}-m_{K}^{2}}f^{-}(s)  \tag{A-4}
\end{equation}

The effective Hamiltonian for the given decay process is given by\cite{SM}

\begin{equation}
H_{eff}=\frac{4G_{F}}{\sqrt{2}}V_{tb}V_{ts}^{\ast }\underset{i=1}{\overset{10%
}{\Sigma }}\{C_{i}(\mu )\ O_{i}(\mu )+C_{Q_{i}}(\mu )\ O_{Q_{i}}(\mu )\}. 
\tag{A-5}
\end{equation}

The Wilson coefficients relevant to our calculaton are evaluated at $\mu
=m_{b}$:

\begin{center}
\begin{equation}
C_{7}=-0.315,\ C_{10}=-4.642.\medskip  \tag{A-6}
\end{equation}
\end{center}

The expression for $C_{9}^{eff}$ in the next to leading order approximation
used here is given by \cite{OP,SM,CB}\medskip

\begin{center}
\begin{equation}
C_{9}^{eff}(\widehat{m}_{b},\widehat{s})=A(\widehat{s})+B\medskip (\widehat{s%
})  \tag{A-7a}
\end{equation}
\end{center}

where

\begin{center}
\begin{equation}
A(\widehat{s})=4.227+0.124w(\widehat{s})+\underset{i=1}{\overset{6}{\Sigma }}%
\alpha _{i}\left( \widehat{s}\right) C_{i}\medskip  \tag{A-7b}
\end{equation}

$\alpha _{1}\left( \widehat{s}\right) =3g(\widehat{m}_{c},\widehat{s});\
\alpha _{2}\left( \widehat{s}\right) =g(\widehat{m}_{c},\widehat{s});\
\alpha _{3}\left( \widehat{s}\right) =3g(\widehat{m}_{c},\widehat{s})-\frac{1%
}{2}g(\widehat{m}_{s},\widehat{s})-2g(\widehat{m}_{b},\widehat{s})+\frac{2}{3%
};\ \alpha _{4}\left( \widehat{s}\right) =\frac{2}{9}+g(\widehat{m}_{c},%
\widehat{s})-\frac{3}{2}g(\widehat{m}_{s},\widehat{s})-2g(\widehat{m}_{b},%
\widehat{s});$

$\alpha _{5}\left( \widehat{s}\right) =\frac{2}{3}+3g(\widehat{m}_{c},%
\widehat{s})-\frac{3}{2}g(\widehat{m}_{b},\widehat{s});\ \alpha _{6}\left( 
\widehat{s}\right) =\frac{2}{9}+g(\widehat{m}_{c},\widehat{s})-\frac{1}{2}g(%
\widehat{m}_{b},\widehat{s})\ \ \ \ \ \ \ \ \ \ \ \ \ \ \ \ \ \ \ \ \ \ \ \
\ \ \ \ \ \ \ \ \ \ \ \ \ \ \ \ \ \ \ \ \ \ \ $
\end{center}

$\ \ \ \ \ \ \ \ \ \ \ \ \ \ \ $%
\begin{equation}
\ \ \ \ B(\widehat{s})=\lambda _{t}(3C_{1}+C_{2})(g(\widehat{m}_{c},\widehat{%
s})-g(\widehat{m}_{u},\widehat{s}))\medskip  \tag{A-7c}
\end{equation}

where coefficients $C_{1}-C_{6}($evaluated at $m_{b})$, $g(\widehat{m}_{q},%
\widehat{s})\ $\ and $w(\widehat{s})$ \cite{OP,SM,CB}\medskip\ used in our
computation are listed as:

\begin{center}
\begin{tabular}{|l|l|l|l|l|l|}
\hline
$C_{1}$ & $C_{2}$ & $C_{3}$ & $C_{4}$ & $C_{5}$ & $C_{6}$ \\ \hline
$-0.249$ & $1.107$ & $0.011$ & $-0.025$ & $0.007$ & $-0.031$ \\ \hline
\end{tabular}
\end{center}

\begin{equation}
g(\widehat{m}_{q},\widehat{s})=\frac{-8}{9}\log [\widehat{m}_{q}]+\frac{4}{9}%
y_{q}-\frac{2}{9}(2+y_{q})\sqrt{1-y_{q}}+\{\theta (1-y_{q})(\log (\frac{1+%
\sqrt{1-y_{q}}}{1-\sqrt{1-y_{q}}})-i\pi )+\theta (y_{q}-1)\tan ^{-1}(\frac{1%
}{\sqrt{y_{q}-1}})\}  \tag{A-7d}
\end{equation}

\begin{equation}
w(s)=-\frac{2}{9}\pi ^{2}-\frac{4}{3}Li_{2}(\widehat{s})-\frac{2}{3}\ln [%
\widehat{s}]\ln (\text{1-}\widehat{s})-\frac{5+4\widehat{s}}{3(+2\widehat{s})%
}\ln (1-\widehat{s})-\frac{2s(1+\widehat{s})(1-2\widehat{s})}{3(1-\widehat{s}%
)^{2}(1+2\widehat{s})}\ln [\widehat{s}]+\frac{5+9\widehat{s}-6(\widehat{s}%
)^{2}}{6(1-\widehat{s})(1+2\widehat{s})}  \tag{A-7e}
\end{equation}

where

$\ \ \ \ \ \ \ \ \ \ \ \lambda _{t}=\frac{V_{ub}^{\ast }V_{us}}{V_{tb}^{\ast
}V_{ts}},\ \widehat{m}_{q}=\frac{m_{q}}{m_{b}},~\widehat{s}=\frac{s}{%
(m_{b})^{2}},$ $y_{q}=\frac{(2\widehat{m}_{q})^{2}}{\widehat{s}}$

$C_{7}$ belongs to photon Penguin and $C_{9}^{eff}$ and $C_{10}$ belong to $%
W $ box and $Z$ Penguin Feynman diagrams contributing to the process under
discussion here. $B$ is continuum part of $u\bar{u}$ and $c\bar{c}$ loops
proportional to $V_{ub}^{\ast }V_{uq}$ and $V_{cb}^{\ast }V_{cq}$
respectively. It is solely responsible for the Wilson contribution to $%
CP-asymmetry.$

\begin{center}
\textbf{Appendix-B}

\textbf{CP-Asymmetry}
\end{center}

We calculate the $A_{CP}$ as

\begin{equation*}
A_{CP}=\frac{\frac{d\Gamma (B\rightarrow Kl^{+}l^{-})}{ds}-\frac{d\Gamma (%
\bar{B}\rightarrow \bar{K}l^{-}l^{+})}{ds}}{\frac{d\Gamma (B\rightarrow
Kl^{+}l^{-})}{ds}+\frac{d\Gamma (\bar{B}\rightarrow \bar{K}l^{-}l^{+})}{ds}}
\end{equation*}%
\begin{eqnarray}
\frac{d\Gamma (\bar{B}\rightarrow \bar{K}l^{-}l^{+})}{ds} &=&\frac{1}{256\pi
^{3}m_{B}^{3}}\beta (m_{l},s)\lambda ^{\frac{1}{2}}(s)\{\left\vert \bar{F}%
_{S}\right\vert ^{2}2s\beta _{l}^{2}+\left\vert \bar{F}_{P}\right\vert
^{2}2s+\left\vert \bar{F}_{V}\right\vert ^{2}\frac{1}{3}\lambda (s)(1+\frac{%
2m_{l}^{2}}{s})  \notag \\
&&+\left\vert \bar{F}_{A}\right\vert ^{2}[\frac{1}{3}\lambda (s)(1+\frac{%
2m_{l}^{2}}{s})+8m_{B}^{2}m_{l}^{2}]+\func{Re}(\bar{F}_{P}^{\ast }\bar{F}%
_{A})4m_{l}(m_{B}^{2}-m_{K}^{2}+s)\},  \TCItag{B-1a}
\end{eqnarray}

\begin{eqnarray}
\frac{d\Gamma (B\rightarrow Kl^{+}l^{-})}{ds} &=&\frac{1}{256\pi
^{3}m_{B}^{3}}\beta (m_{l},s)\lambda ^{\frac{1}{2}}(s)\{\left\vert
F_{S}\right\vert ^{2}2s\beta _{l}^{2}+\left\vert F_{P}\right\vert
^{2}2s+\left\vert F_{V}\right\vert ^{2}\frac{1}{3}\lambda (s)(1+\frac{%
2m_{l}^{2}}{s})  \notag \\
&&+\left\vert F_{A}\right\vert ^{2}[\frac{1}{3}\lambda (s)(1+\frac{2m_{l}^{2}%
}{s})+8m_{B}^{2}m_{l}^{2}]+\func{Re}(F_{P}^{\ast
}F_{A})4m_{l}(m_{B}^{2}-m_{K}^{2}+s)\},  \TCItag{B-1b}
\end{eqnarray}

\begin{equation*}
A_{CP}=\lambda (s)(1+\frac{2m_{l}^{2}}{s})(\left\vert F_{V}\right\vert
^{2}-\left\vert \bar{F}_{V}\right\vert ^{2})(D(s))^{-1}
\end{equation*}

where

\begin{eqnarray}
D(s) &=&6(\left\vert F_{S}\right\vert ^{2}2s\beta (m_{l},s)^{2}+\left\vert
F_{P}\right\vert ^{2}2s+\left\vert F_{A}\right\vert ^{2}[\frac{1}{3}\lambda
(s)(1+\frac{2m_{l}^{2}}{s})+8m_{B}^{2}m]+  \TCItag{B-2} \\
&&\func{Re}(F_{P}F_{A}^{\ast })4m_{l}(m_{B}^{2}-m_{K}^{2}+s)+\frac{\lambda
(s)}{6}(1+\frac{2m_{l}^{2}}{s})(\left\vert F_{V}\right\vert ^{2}+\left\vert 
\bar{F}_{V}\right\vert ^{2}))  \notag
\end{eqnarray}

where

\begin{eqnarray}
F_{V} &=&\frac{G_{F}\alpha V_{tb}V_{ts}^{\ast }}{2\sqrt{2}\pi }(2C_{9}^{eff}(%
\widehat{m}_{b},\widehat{s})f^{+}(s)-C_{7}\frac{4m_{b}}{m_{B}+m_{K}}%
f^{T}(s))+\frac{1}{4}f^{+}(s)\underset{i,m,n=1}{\overset{3}{\sum }}%
V_{ni}^{\dagger }V_{im}\frac{\lambda _{\beta n3}^{\prime }\lambda _{\beta
m2}^{\prime \ast }}{m_{\widetilde{u_{i}^{c}}}^{2}},  \TCItag{B-3} \\
F_{A} &=&\frac{G_{F}\alpha V_{tb}V_{ts}^{\ast }}{2\sqrt{2}\pi }%
(2C_{10}f^{+}(s))-\frac{1}{4}f^{+}(s)\overset{3}{\underset{i,m,n=1}{\sum }}%
V_{ni}^{\dagger }V_{im}\frac{\lambda _{\beta n3}^{\prime }\lambda _{\beta
m2}^{\prime \ast }}{m_{\widetilde{u_{i}^{c}}}^{2}},  \notag \\
F_{S} &=&\frac{1}{4}\frac{(m_{B}^{2}-m_{K}^{2})}{m_{b}-m_{s}}f_{o}(s)%
\underset{i=1}{\overset{3}{\sum }}\frac{(\lambda _{i\beta \beta }^{\ast
}\lambda _{i23}^{\prime }-\lambda _{i\beta \beta }\lambda _{i32}^{\prime
\ast })}{m_{\widetilde{\nu }_{Li}}^{2}},  \notag \\
F_{P} &=&\frac{G_{F}\alpha V_{tb}V_{ts}^{\ast }}{2\sqrt{2}\pi }%
2m_{l}C_{10}(f^{+}(s)+f^{-}(s))+\frac{1}{4}\frac{(m_{B}^{2}-m_{K}^{2})}{%
m_{b}-m_{s}}f_{o}(s)\underset{i=1}{\overset{3}{\sum }}\frac{(\lambda
_{i\beta \beta }^{\ast }\lambda _{i23}^{\prime }+\lambda _{i\beta \beta
}\lambda _{i32}^{\prime \ast })}{m_{\widetilde{\nu }_{Li}}^{2}},  \notag
\end{eqnarray}

\begin{eqnarray}
\bar{F}_{V} &=&\frac{G_{F}\alpha V_{tb}V_{ts}^{\ast }}{2\sqrt{2}\pi }(2\bar{C%
}_{9}^{eff}f^{+}(s)-C_{7}\frac{4m_{b}}{m_{B}+m_{K}}f^{T}(s))+\frac{1}{4}%
\underset{m,n,i=1}{f^{+}(s)(\overset{3}{\sum }}V_{ni}^{\dagger }V_{im}\frac{%
\lambda _{\beta n3}^{\prime }\lambda _{\beta m2}^{\prime \ast }}{m_{%
\widetilde{u_{i}^{c}}}^{2}})^{\ast },  \TCItag{B-4} \\
\bar{F}_{A} &=&F_{A}^{\ast },\ \bar{F}_{S}=F_{S}^{\ast },\ \bar{F}%
_{P}=F_{p}^{\ast },\bar{C}_{9}^{eff}(\widehat{m}_{b},\widehat{s}%
)=C_{9}^{eff}(\widehat{m}_{b},\widehat{s},\lambda _{t}^{\ast }),  \notag \\
\lambda _{\beta n3}^{\prime }\lambda _{\beta m2}^{\prime \ast }
&=&\left\vert \lambda _{\beta n3}^{\prime }\lambda _{\beta m2}^{\prime \ast
}\right\vert e^{i\Theta },\beta =e,\mu  \TCItag{B-5}
\end{eqnarray}%
\medskip

Eq. (18) for $A_{CP}$ contains the following factor

\begin{eqnarray}
\left\vert F_{V}\right\vert ^{2}-\left\vert \bar{F}_{V}\right\vert ^{2} &=&4(%
\frac{G_{F}\alpha V_{tb}V_{ts}^{\ast }}{2\sqrt{2}\pi })^{2}f^{+}(s)\func{Im}%
(A^{\prime }(s)B^{\prime }(s)^{\ast })\func{Im}(\lambda _{t})+  \notag \\
&&\frac{G_{F}\alpha V_{tb}V_{ts}^{\ast }}{2\sqrt{2}\pi }f^{+}(s)\func{Im}%
(A^{\prime }(s))\func{Im}(\underset{m,n,i=1}{\overset{3}{\sum }}%
V_{ni}^{\dagger }V_{im}\frac{\lambda _{\beta n3}^{\prime }\lambda _{\beta
m2}^{\prime \ast }}{2m_{\widetilde{u_{i}^{c}}}^{2}})+  \notag \\
&&\frac{G_{F}\alpha V_{tb}V_{ts}^{\ast }}{2\sqrt{2}\pi }(f^{+}(s))^{2}\func{%
Im}(B^{\prime }(s))\func{Im}(\underset{m,n,i=1}{\overset{3}{\lambda
_{t}^{\ast }\sum }}V_{ni}^{\dagger }V_{im}\frac{\lambda _{\beta n3}^{\prime
}\lambda _{\beta m2}^{\prime \ast }}{2m_{\widetilde{u_{i}^{c}}}^{2}}), 
\TCItag{B-6}
\end{eqnarray}

where

\begin{eqnarray*}
A^{\prime }(s) &=&2Af^{+}(s)-C_{7}\frac{4m_{b}}{m_{B}+m_{K}}f^{T}(s) \\
B^{\prime }(s) &=&2(3C_{1}+C_{2})(g(\widehat{m}_{c},\widehat{s})-g(\widehat{m%
}_{u},\widehat{s})),
\end{eqnarray*}

\begin{center}
\begin{eqnarray}
\left\vert F_{V}\right\vert ^{2}+\left\vert \bar{F}_{V}\right\vert ^{2}
&=&G(s)+H(s),  \notag \\
G(s) &=&2(f^{+}(s))^{2}(\left\vert \underset{m,n,i=1}{\overset{3}{\sum }}%
V_{ni}^{\dagger }V_{im}\frac{\lambda _{\beta n3}^{\prime }\lambda _{\beta
m2}^{\prime \ast }}{m_{\widetilde{u_{i}^{c}}}^{2}}\right\vert )^{2}+4(\frac{%
G_{F}\alpha V_{tb}V_{ts}^{\ast }}{2\sqrt{2}\pi })f^{+}(s)\func{Re}(A^{\prime
})  \notag \\
&&\func{Re}(\underset{m,n,i=1}{\overset{3}{\sum }}V_{ni}^{\dagger }V_{im}%
\frac{\lambda _{\beta n3}^{\prime }\lambda _{\beta m2}^{\prime \ast }}{2m_{%
\widetilde{u_{i}^{c}}}^{2}})+4\frac{G_{F}\alpha V_{tb}V_{ts}^{\ast }}{2\sqrt{%
2}\pi }(f^{+}(s))^{2}\func{Re}(\lambda _{t}^{\ast }\overset{3}{\underset{%
i,m,n=1}{\sum }}V_{ni}^{\dagger }V_{im}\frac{\lambda _{\beta n3}^{\prime
}\lambda _{\beta m2}^{\prime \ast }}{2m_{\widetilde{u_{i}^{c}}}^{2}})\func{Re%
}(B^{\prime }),  \TCItag{B-7} \\
H(s) &=&(\frac{G_{F}\alpha V_{tb}V_{ts}^{\ast }}{2\sqrt{2}\pi }%
)^{2}(\left\vert 2C_{9}^{eff}(\widehat{m}_{b},\widehat{s})f^{+}(s)-C_{7}%
\frac{4m_{b}}{m_{B}+m_{K}}f^{T}(s)\right\vert ^{2}+\left\vert 2\bar{C}%
_{9}^{eff}(\widehat{m}_{b},\widehat{s})f^{+}(s)-C_{7}\frac{4m_{b}}{%
m_{B}+m_{K}}f^{T}(s)\right\vert ^{2})  \notag
\end{eqnarray}
\end{center}

The factors contributing to the CP- asymmetry at various energies are;

\begin{eqnarray*}
\func{Im}(A^{\prime }(s)) &=&2f^{+}(s)\func{Im}(A(s)=2f^{+}(s)\left\vert 
\begin{array}{c}
\frac{1}{2}\pi (C_{3}+3C_{4}) \\ 
-\pi (3C_{1}+C_{2}+3C_{3}+C_{4}+3C_{5}+C_{6})+\frac{1}{2}\pi (C_{3}+3C_{4})
\\ 
-\pi (3C_{1}+C_{2}+3C_{3}+C_{4}+3C_{5}+C_{6})+\frac{1}{2}\pi (C_{3}+3C_{4})+%
\frac{1}{2}\pi (4C_{3}+4C_{4}+3C_{5})%
\end{array}%
\right\} 
\begin{array}{c}
m_{c}^{2}>s>m_{s}^{2} \\ 
m_{b}^{2}>s>m_{c}^{2} \\ 
s>m_{b}^{2}%
\end{array}
\\
\func{Im}(B^{\prime }(s)) &=&2(3C_{1}+C_{2})\func{Im}(g(\widehat{m}_{c},%
\widehat{s})-g(\widehat{m}_{u},\widehat{s})), \\
&=&2(3C_{1}+C_{2})\left\vert 
\begin{array}{c}
\frac{2\pi }{9}(2+y_{u})\sqrt{1-y_{u}} \\ 
\frac{2\pi }{9}((2+y_{u})\sqrt{1-y_{u}}-(2+y_{c})\sqrt{1-y_{c}})%
\end{array}%
\right\} 
\begin{array}{c}
m_{c}^{2}>s>m_{u}^{2} \\ 
s>m_{c}^{2}%
\end{array}%
\end{eqnarray*}%
\newline
\begin{equation*}
\lambda _{t}=-C\lambda ^{2}e^{i\delta }
\end{equation*}

Since for simplicity, we have considered only the contribution from $i=3$ in 
$\NEG{R}_{p}$ Yukawa couplings, $V_{tb}$ contributes in $F_{A,V}$ only.
Further, since ($V_{td},V_{ts})\symbol{126}10^{-3},$ we neglect these
numbers and take

$V_{tb}\symbol{126}1\cite{pdg}.$\ Since$\cite{CP,pdg}$

\begin{equation*}
0.23<C<0.59;\ 0.216<\lambda _{t}<0.223;\ \delta =1.3439
\end{equation*}

we use the central value of $\lambda _{t}$ and $C$,i.e. 0.41and 0.22
respectively. This parameterises the CP-violating phase used in Fig 10.

\end{document}